\documentclass[manuscript]{acmart}

\AtBeginDocument{%
  \providecommand\BibTeX{{%
    \normalfont B\kern-0.5em{\scshape i\kern-0.25em b}\kern-0.8em\TeX}}}

\copyrightyear{2023}
\acmYear{2023}
\setcopyright{rightsretained}
\acmConference[RecSys '23]{Seventeenth ACM Conference on Recommender Systems}{September 18--22, 2023}{Singapore, Singapore} \acmBooktitle{Seventeenth ACM Conference on Recommender Systems (RecSys '23), September 18--22, 2023, Singapore, Singapore}\acmDOI{10.1145/3604915.3608870} \acmISBN{979-8-4007-0241-9/23/09}

\newif\ifproofread

\usepackage{color,soul}
\usepackage{pgfplots}
\usepackage{hyperref}
\usepackage{framed}
\usepackage{fontawesome}

\usepackage{bm}
\usepackage{multirow}
\usepackage{makecell}
\usepackage{booktabs} 
\usepackage{multirow,bigdelim}
\usepackage{easybmat}
\usepackage{threeparttable}
\usepackage{pgfplotstable}
\pgfplotsset{compat=newest}
\usepackage{caption}
\usepackage{subcaption}
\usepackage{tikz}

\usepackage{xspace}
\usepackage{xcolor}
\usepackage{xargs} 
\usepackage[colorinlistoftodos,prependcaption,textsize=tiny]{todonotes}

\newcommandx{\add}[2][1=]{\todo[linecolor=red,backgroundcolor=red!25,bordercolor=red,#1]{#2}}
\newcommandx{\change}[2][1=]{\todo[linecolor=blue,backgroundcolor=blue!25,bordercolor=blue,#1]{#2}}
\newcommandx{\info}[2][1=]{\todo[linecolor=OliveGreen,backgroundcolor=OliveGreen!25,bordercolor=OliveGreen,#1]{#2}}
\newcommandx{\improvement}[2][1=]{\todo[linecolor=orange,backgroundcolor=orange!25,bordercolor=orange,#1]{#2}}

\usepackage{listings}
\usepackage{xcolor}

\definecolor{codegreen}{rgb}{0,0.6,0}
\definecolor{codegray}{rgb}{0.5,0.5,0.5}
\definecolor{codepurple}{rgb}{0.58,0,0.82}
\definecolor{backcolour}{rgb}{0.95,0.95,0.92}

\lstdefinestyle{mystyle}{
    backgroundcolor=\color{backcolour},   
    commentstyle=\color{codegreen},
    keywordstyle=\color{magenta},
    numberstyle=\tiny\color{codegray},
    stringstyle=\color{codepurple},
    basicstyle=\ttfamily\footnotesize,
    breakatwhitespace=false,         
    breaklines=false,                 
    captionpos=b,                    
    keepspaces=false,                 
    numbers=left,                    
    numbersep=4pt,                  
    showspaces=false,                
    showstringspaces=false,
    showtabs=false,                  
    tabsize=1
}

\usepackage{float}
\usepackage{amsmath,amsfonts,bm}

\def\eqref#1{equation~\ref{#1}}
\def\1{\bm{1}}

\DeclareMathAlphabet{\mathsfit}{\encodingdefault}{\sfdefault}{m}{sl}
\SetMathAlphabet{\mathsfit}{bold}{\encodingdefault}{\sfdefault}{bx}{n}

\usepackage{soul}
\begin{document}

%%
%% The "title" command has an optional parameter,
%% allowing the author to define a "short title" to be used in page headers.
\title{AdaptEx: a self-service contextual bandit platform}

%%
%% The "author" command and its associated commands are used to define
%% the authors and their affiliations.
%% Of note is the shared affiliation of the first two authors, and the
%% "authornote" and "authornotemark" commands
%% used to denote shared contribution to the research.

 \author{William Black}
 \authornotemark[1]
 \author{Ercument Ilhan}
 \authornotemark[1]
 \author{Andrea Marchini}
 \authornotemark[1]
 \author{Vilda Markeviciute}
\affiliation{Expedia Group}
\renewcommand{\shortauthors}{W. Black, E. Ilhan, A. Marchini and  V. Markeviciute}

\authornote{All the authors contributed equally to this paper.}

%%
%% The abstract is a short summary of the work to be presented in the
%% article.
\begin{abstract}
This paper presents AdaptEx, a self-service contextual bandit platform widely used at Expedia Group, that leverages multi-armed bandit algorithms to personalize user experiences at scale. AdaptEx considers the unique context of each visitor to select the optimal variants and learns quickly from every interaction they make. It offers a powerful solution to improve user experiences while minimizing the costs and time associated with traditional testing methods. The platform unlocks the ability to iterate towards optimal product solutions quickly, even in ever-changing content and continuous "cold start" situations gracefully.
\end{abstract}

%%
%% The code below is generated by the tool at http://dl.acm.org/ccs.cfm.
%% Please copy and paste the code instead of the example below.
%%



\maketitle

\section{Introduction}\label{sec:introduction}
Expedia Group is a technology company powering travel that has revolutionised the way people search for and book travel through its B2B network and house of brands.
In particular, Expedia Group relies on their apps to generate bookings and revenue.
By optimizing the presentation of information and options on the apps, Expedia Group can make it easier and more compelling for users to make a booking or purchase. 
Contextual multi-armed bandits (MAB) \cite{lattimore2020bandit} have emerged as a powerful tool for optimizing user experiences in a variety of industries \cite{chapelle2014simple, bouneffouf2020survey, amat2018artwork}.
With the rise of big data and artificial intelligence, companies have more information at their disposal than ever before, and it is becoming increasingly important to quickly iterate towards optimal product solutions that meet the unique needs and preferences of each individual user.
The MAB problem involves a decision-maker selecting from a set of options, each of which provides a reward.
In the absence of prior knowledge about the reward distribution, the decision-maker must balance the desire to learn more about the options (exploration) with the desire to select the option that has the highest expected reward (exploitation).
A contextual MAB explores different variants by considering also the unique context of each user and selecting the variant that is most likely to lead to a positive outcome.
Over time, the bandit learns from every interaction and adapts its selection strategy to maximize the rewards.

AdaptEx is a self-service contextual bandit platform that is designed to address the challenge of setting up bandits algorithms to select the best user experience. 
AdaptEx's self-service approach empowers any product team to quickly and easily configure and deploy contextual bandits without requiring specialised machine learning expertise.
The platform's intuitive API make it accessible to anyone who wants to improve their product's user experience.
Furthermore, the self-service model allows clients to maintain full control over the bandit's configuration and operation, enabling them to tailor the solution to their specific needs and goals.
Traditional testing methods can be very time consuming and require a large number of interactions with each variant to reach statistical significance.
Meanwhile, AdaptEx uses contextual MAB to adapt to user behaviour and gradually discard the suboptimal solutions from a much larger set of options allowing faster product iteration.
The AdaptEx platform requires no prior model training and is designed to be flexible and scalable.
The platform also handles ever-changing content and continuous "cold start" situations gracefully, allowing users to quickly iterate towards optimal solutions without being hindered by data limitations.

In this paper we present the architecture, algorithms and use cases served by AdaptEx as well as discuss future challenges.

\section{AdaptEx Platform Architecture}\label{sec:architecture}
\begin{figure}[htp]
    \centering
    \includegraphics[width=0.8\columnwidth]{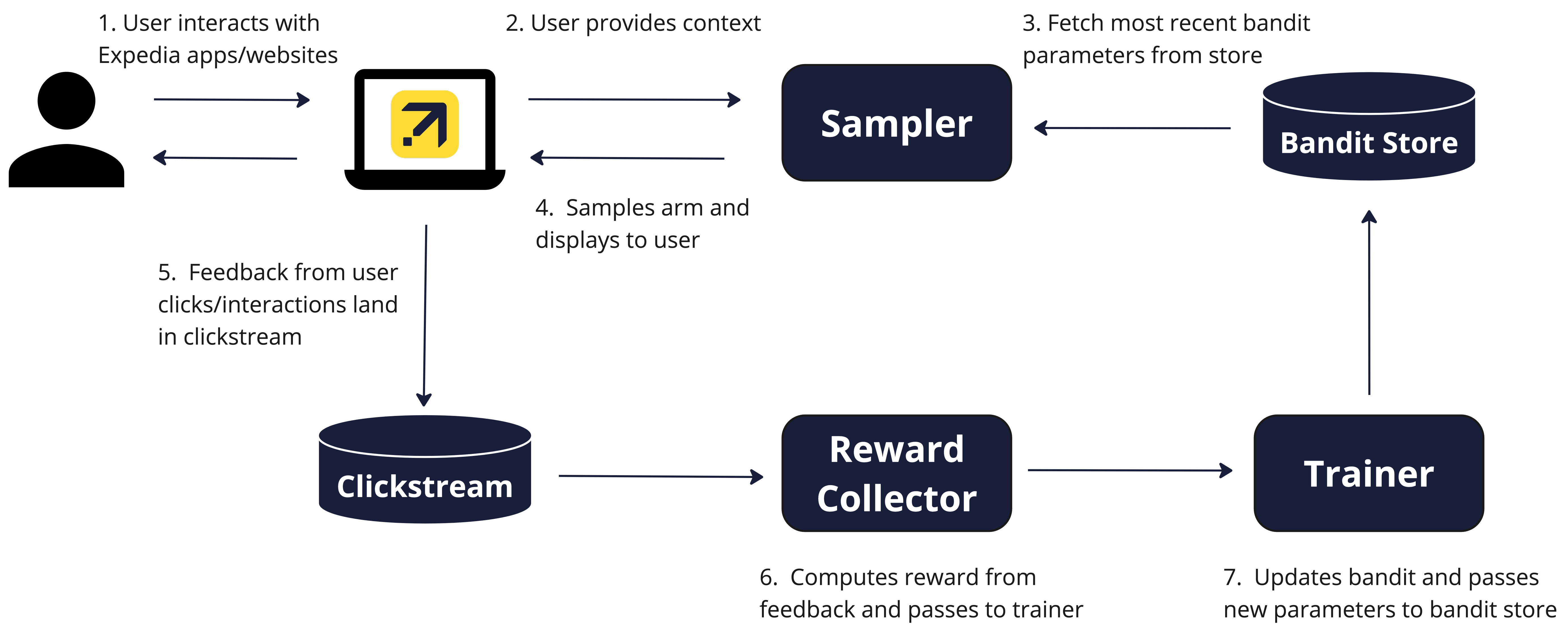}
    \Description[AdaptEx Platform Architecture Overview]{A step by step display of how the AdaptEx Platform learns from customers interactions in real time. 1: User interacts with Expedia apps/websites. 2: User provides context. 3: Fetch most recent bandit parameters from store. 4: Samples arm and displays to user. 5: Feedback from user clicks/interactions land in clickstream. 6: Computes reward from feedback and passes to trainer. 7: Updates bandit and passes new parameters to bandit store.}
    \caption{AdaptEx Platform Architecture Overview}
    \label{fig:Architecture}
\end{figure}

There are several components that make up AdaptEx, as shown in Figure \ref{fig:Architecture}, which enables the platform to adaptively select the optimal experience to show to the user, while learning in real-time from their implicit feedback.

When a user visits a page, they can provide their context explicitly, for example by entering a holiday destination or travel dates, or implicitly through features such as their device type or time of day.
In the AdaptEx platform, this information is passed as a request to the Sampler who's role is to return an "arm" (an experience) to the user according to the bandit's recommendation.
To do this, the Sampler periodically fetches the most up-to-date parameters from the Bandit Store, which is a service backed by a MongoDB database that stores all the bandits' parameters and configurations.
It then samples an arm according to the configured sampling algorithm and returns this experience back to the user (see Section \ref{sec:algos} for a review of algorithms used). 
The Sampler is an auto-scaling Spring Boot app, designed to scale horizontally according to traffic, ensuring that the platform can dynamically handle high levels of user activity, which vary greatly due to seasonal trends, without affecting performance. 
Additionally, the Sampler includes a consistency cache to ensure the user experience remains consistent across a session.

As the user interacts with the application/website, their actions are logged in the clickstream via a Kafka \cite{kreps2014kafka} topic. 
The Reward Collector then extracts the interactions that are relevant to the bandit and aggregates them into rewards using a Flink \cite{carbone2015apache} job.

The rewards are then passed to the Trainer in batches through another Kafka topic. 
Then, via a Spark Job \cite{10.5555/1863103.1863113}, the Trainer updates the bandit's parameters according to the configured algorithm that fits the use case.
The updated parameters are then passed to the Bandit Store, completing the feedback loop.

The AdaptEx platform is designed to be self-serve, allowing clients to interact with the platform through a simple API call, using Swagger UI, to configure the bandit with suitable arms, contexts, and rewards.
This lands the configuration in the Bandit Store, instantiating a model which is ready to learn.
The clients then only need to set up tagging for the relevant events, such as user contexts, arms displayed and reward events like clicks, which allows the Reward Collector to collect them from the clickstream and pass them to back the bandit.

In this way no new models need to be deployed by clients and they can instantiate a bandit in an off-the-shelf manner, enabling rapid development and deployment, expediting the optimisation of the site and roll out of new adaptive features.
\section{Algorithms Implemented in AdaptEx}\label{sec:algos}
AdaptEx employs a variety of algorithms, both classic and contextual, to effectively optimize the selection of experiences.% for each user.

The classic MAB algorithms that are used are Epsilon Greedy (EG) \cite{sutton2018reinforcement}, Thompson Sampling (TS) \cite{agrawal2012analysis, russo2018tutorial, thompson1933likelihood}, and Exponential-weight algorithm for Exploration and Exploitation \cite{auer2002nonstochastic}.
These algorithms are chosen because they are robust to delayed feedback.
TS is the most commonly used algorithm on the platform, which maintains a posterior distribution over the reward distributions of each arm.
It samples a value from each arm's posterior distribution at each time step and selects the arm with the highest sampled value.
As the algorithm observes more data, the posterior distributions become more concentrated, and the algorithm converges to the optimal arm. 

AdaptEx uses several linear contextual bandit algorithms that use a linear model to estimate the expected reward of each action based on contextual variables \cite{Parfenov2021}. 
The commonly used methods to update the model are Bayesian Logistic Regression (BLR) \cite{gelman1995bayesian} and Recursive Least Squares (RLS) \cite{yin2003stochastic}. 
BLR is often used for binary rewards, where a logistic regression model is updated using Bayesian inference to estimate the probability of receiving a positive reward given the context and the chosen arm.
RLS uses an online linear regression model to estimate the expected reward given the context and the chosen action for continuous rewards.
In addition to TS and EG for contextual bandits \cite{chapelle2011empirical}, Inverse Gap Weighting (IGW) \cite{foster2021efficient} is another algorithm AdaptEx uses.
It selects arms with probabilities proportional to the gaps between their expected rewards.
Cascading Bandits is used when the bandit must rank items.
It employs TS to select a sequence of items to display to the user, where each item is chosen based on the probability that it will be clicked given the previous item displayed \cite{zong2016cascading, zhong2021thompson}.
When the bandit must balance multiple objectives, AdaptEx uses the Generalized Gini Index aggregation function to scalarize each objective's rewards to find a solution on the Pareto Front \cite{mehrotra2020bandit}.
Finally, in addition to solving bandit problems, AdaptEx also needs to be computationally efficient to cope with the low latency requirements of real-world applications.
Thus, AdaptEx incorporates Greedy Search \cite{Parfenov2021} in the best-arm determination process of every arm sampling request in latency-prone applications.
Instead of exhaustively iterating through every arm option, this algorithm finds either the exact or a good quality approximation solution while respecting the time constraints to respond these requests.

\section{AdaptEx Use Cases}\label{sec:use_cases}
The self-serve nature of the AdaptEx platform ensures that any product team can easily configure a MAB. 
Teams can choose the configuration of arms and contextual features which best suit their hypotheses, and also set the bandit's rewards to align with a variety of business metrics including customer engagement, loyalty and bookings.
The life-cycle of learning and testing the bandit-decided experiences is as follows.
First, the bandit enters the learning phase during which it explores different experiences and learns from the received user feedback.
Then, the learning phase is stopped by fixing the best variants (‘arms’) determined by the bandit: the bandit stops exploring and starts fully exploiting its learnings. This frozen bandit is then tested against the control experience in an A/B test to ensure the changes are statistically significant against chosen business metrics.

Such platform flexibility and compatibility with the existing experimentation methodology have attracted a wide range of use cases of varying complexity from different Expedia Group brands and business functions.

A simple non-contextual example is content selection for a website module from a set of ${\sim}10$ candidates.
Such a MAB test can fine-tune the user experience and provide content clarity, by finding the winning variant in as quickly as one week without the need to run multiple A/B tests which would take about ${\sim}10$ times longer. 

A more typical configuration involves the optimisation of multiple module components on a page, by personalising their content, design and placement to the user giving ${\sim}100$ different combinations to choose from. 
The bandit would make a highly-personalised choice for over 1000 different customer segments based on features such as trip type or point of sale, totalling to over 100 000 arm-context combinations.  
Such MAB tests could take 2-4 weeks to find the best choice for each customer segment, depending on the traffic exposure. Such fast personalisation from a large number of possible variants would be impossible with a traditional A/B test.

The platform architecture and sophisticated algorithms have allowed AdaptEx to solve even more highly-dimensional use cases.
Some examples include choosing the best possible main (“hero”) image for each property\footnote{\text{See more:\, }\mbox{\url{https://medium.com/expedia-group-tech/how-we-optimized-hero-images-on-hotels-com-using-multi-armed-bandit-algorithms-4503c2c32eae}}}, or ranking dozens of items or modules on a page, which result in millions of possible permutations.

Finally, AdaptEx can help other Machine Learning (ML) teams overcome cold start problems. For example, it can determine the best recommendation strategy for a new customer when there is no historical data available, or find the optimal product placement for a new item when there is no data on customer behavior. By providing context-specific recommendations and optimizing product placement for new items, AdaptEx can assist other ML teams collecting data and learning in real time.

Since launching, AdaptEx has shown promising results in learning a personalized experience that very often performs better than the control group.

The success of the platform lies within the MAB algorithms’ ability to test a large selection of arms quickly discarding the obviously sub-optimal options. 
The highly-customisable self-serve format of the platform has enabled a wide range of teams with varying business needs to take advantage of the capabilities allowing faster personalised feature testing and development.
%\section{Conclusion and Future Work}

%In this paper, we presented Expedia Group's self-service contextual bandit platform named AdaptEx.
%Despite its wide range of solutions, there are still some capabilities that fall out of its scope.

\section{Future Work}

Despite the wide range of solutions AdaptEx offers, there are still some capabilities that fall out of its scope.
We plan to develop AdaptEx further in several ways to extend its solution coverage.

\textbf{Non-linear approaches:} 
Even though linear models can be advantageous at learning and inference time with proven performance guarantees \cite{agrawal2013thompson}, their limited representational power may be insufficient to capture complex reward dynamics.
%Others in the field have shared this motivation, which has driven some of the most recent advances \cite{DBLP:conf/icml/Zhou0G20, DBLP:conf/iclr/0002WZG22, zhang2021neural}.
We will look to employ non-linear algorithms, such as neural networks, to keep the performance of AdaptEx up-to-date with cutting-edge solutions and drive further personalisation even in the most complex reward settings.

\textbf{Reinforcement Learning (RL):}
Not every user interaction can be expected to be limited to a single step.
It is common for them to be part of a multi-step experience such as the journey from landing on the home page through to a successful booking.
However, MAB by default are not designed to model and solve such problems despite their relevancy and importance to us.
Thus, we consider leveraging RL techniques \cite{RL} %\cite{DBLP:conf/aaai/HesselMHSODHPAS18, DBLP:conf/icml/HaarnojaZAL18}
as an addition to AdaptEx to handle use cases that are better modelled as sequential decision-making problems.

%\begin{itemize}
%\item \textbf{Non-linear Bandits:} 
%Even though linear models can be advantageous at learning and inference time with proven performance guarantees \cite{agrawal2013thompson}, their limited representational power may be insufficient to capture complex reward dynamics.
%For this reason we will look to employ non-linear methods such as neural networks to tackle such problems.
%Others in the field have shared this motivation, which has driven some of the most recent advances \cite{DBLP:conf/icml/Zhou0G20, DBLP:conf/iclr/0002WZG22, zhang2021neural}.
%One of our next steps is to adopt this class of algorithms in our platform to keep its performance up-to-date with cutting-edge solutions and drive further personalisation even in the most complex reward settings.

%\item \textbf{Reinforcement Learning:}
%Not every user interaction can be expected to be limited to a single step.
%It is common for them to be part of a multi-step experience such as the journey from landing on the home page through to a successful booking.
%However, MAB by default are not designed to model and solve such problems despite their relevancy and importance to us.
%Thus, we consider leveraging prominent Reinforcement Learning techniques \cite{DBLP:conf/aaai/HesselMHSODHPAS18, DBLP:conf/icml/HaarnojaZAL18} as a vital addition to AdaptEx to handle use cases that are better modelled as sequential decision-making problems.
%\end{itemize}
% \input{SECTIONS/bios}

% Introduction (Brief literature review will be here)
% AdaptEx
% Use Cases

% Future Work
\section*{Acknowledgements}
We would like to thank the following people for their input to AdaptEx: Gyula Magyar, Vasilis Manolis, Agnes Rozsas, and Balazs Varkoly (engineering team); Remi Diana and Wen Wong (team leaders); Ludovik Çoba for his valuable feedback on the paper; and all previous members of the Reinforcement Learning team.

\bibliographystyle{ACM-Reference-Format}
\bibliography{biblio}

\end{document}